# Transcranial low-level laser stimulation in near infrared-II region for brain safety and protection

Zhilin Li[1,2#], Yongheng Zhao[3#], Yiqing Hu[2,4], Yang Li[2], Keyao Zhang[5], Zhibing Gao[6], Lirou Tan[2,7],

Hanli Liu[4], Xiaoli Li[8], Aihua Cao[3*], Zaixu Cui[1,2*], Chenguang Zhao[2*]

[1]School of Basic Medical Sciences, Capital Medical University, Beijing 100069, China

[2]Chinese Institute for Brain Research, Beijing 102206, China

[3]Department of Pediatrics, Qilu Hospital of Shandong University, Jinan 250012, Shandong Province, China

[4]Department of Bioengineering, the University of Texas at Arlington, Arlington, Texas 76019, USA

[5]School of Biomedical Engineering, Faculty of Medicine, Dalian University of Technology, Dalian 116081, China

[6]National Center for Mental Health, Beijing 100029, China

[7]College of Biological Sciences, China Agricultural University, Beijing 100091, China

[8]School of Automation Science and Engineering, South China University of Technology, Guangzhou 510641, China

[#]These authors contributed equally to this work

[*]Corresponding Author：

Chenguang Zhao; Email: chenguang918.zhao@gmail.com

Zaixu Cui; Email: cuizaixu@cibr.ac.cn

Aihua Cao; Email: xinercah@163.com

# Abstract

**Background:** The use of near-infrared lasers for transcranial photobiomodulation (tPBM) offers a non-invasive method for influencing brain activity and is beneficial for various neurological conditions.

**Objective:** To investigate the safety and neuroprotective properties of tPBM using near-infrared (NIR)-II laser stimulation.

**Methods:** We conducted thirteen experiments involving multidimensional and quantitative methods and measured serum neurobiomarkers, performed electroencephalogram (EEG) and magnetic resonance imaging (MRI) scans, assessed executive functions, and collected a subjective questionnaire.

**Results:** Significant reductions in neuron-specific enolase (NSE) levels were observed after treatment, indicating neuroprotective effects. No structural or functional brain abnormalities were observed, confirming the safety of tPBM. Additionally, cognitive and executive functions were not impaired, with participants' feedback indicating minimal discomfort.

**Conclusions:** Our data indicate that NIR-II tPBM is safe within specific parameters, highlighting its potential for brain protection.

# Introduction

Albert Einstein first proposed the theory of stimulated emission in 1917, which led to the development of lasers. Lasers have since been widely used in various clinical applications, including surgical treatment, dermatological therapy, and photodynamic therapy (Cotler *et al.* 2015; Hashmi *et al.* 2010; Posten *et al.* 2005; Rojas and Gonzalez-Lima 2011). When the laser energy is reduced for application to the brain, the technology is referred to as transcranial photobiomodulation (tPBM). As an emerging non-invasive neuromodulation technology, tPBM utilizes low-level near-infrared (NIR) light wavelengths in NIR-I (760–900 nm) and NIR-II (1000–1700 nm) (Lin *et al.* 2024). NIR light can penetrate the skin and skull to modulate the cortical activity (Salehpour *et al.* 2018). At the cellular level, NIR light activates cytochrome c oxidase (CCO), a specific enzyme in the mitochondrial electron transport chain that serves as the primary photoreceptor responsible for the tPBM effects on the brain (Rojas and Gonzalez-Lima 2011). This activation enhances adenosine triphosphate (ATP) production (Mochizuki-Oda *et al.* 2002), efficiently scavenges reactive oxygen species (Huang *et al.* 2013; Liang *et al.* 2006), and boosts cerebral blood flow (Uozumi *et al.* 2010).

Over the past two decades, extensive research on animal models has driven the use of tPBM in human cognition (Lee *et al.* 2023) and disease research (Gutiérrez-Menéndez *et al.* 2020). Similar to the animal studies (Lapchak *et al.* 2004; Oron *et al.* 2007), human studies have demonstrated tPBM's ability to modulate brain activity (Dole *et al.* 2023), enhance cognition (Zhao *et al.* 2022), and improve clinical performance (Lin *et al.* 2024). Initially, many human tPBM studies (Caldieraro and Cassano 2019; Hacke *et al.* 2014; Huisa *et al.* 2013; Lampl *et al.* 2007; Zivin *et al.* 2009a) focused on the effectiveness and safety of NIR-I wavelengths. These wavelengths were chosen because of their maximum absorption by CCO, which helps to optimize the regulation of cellular metabolism (Li *et al.* 2018). In the last five years, there has been growing interest in human tPBM research in the NIR-II region (Lin *et al.* 2024; Penberthy and Vorwaller 2021; Zhao *et al.* 2022).

Compared with NIR-I, tPBM in NIR-II exhibited lower scattering rates, suggesting its potential to stimulate deeper brain regions (Salehpour et al. 2018) and elicit wavelength-specific effects. For example, recent research has found that, in comparison to 852 nm light, 1064 nm tPBM applied to the right prefrontal cortex can enhance working memory (Zhao et al. 2022). In addition to its specific effects, NIR-II can also perform the neural modulation effects of NIR-I. For example, 1064 nm tPBM applied to the right human forehead has been shown to modify cerebral hemodynamics, metabolism, electroencephalogram (EEG) oscillation power, and functional connectivity (Pruitt et al. 2020; Shahdadian et al. 2022; Truong et al. 2022). Furthermore, it has been found 1064 nm tPBM repeated for 7 days can enhance the working memory of healthy older adults, with its effectiveness lasting for at least 3 weeks (Qu et al. 2022). Collectively, these results highlighted the potential importance of NIR-II tPBM.

Current safety studies on tPBM primarily utilize LED devices in the NIR-I region to assess potential adverse effects in a large cohort of participants (Cassano et al. 2022; Zivin et al. 2009b). Unlike LEDs, lasers are capable of delivering large amounts of energy in very brief durations and over small areas because of their temporal and spatial coherence. This contributes to their effectiveness, but also poses potential risks when interacting with brain tissue (Hecht 2010; Hohberger 2016). Although placing a beamformer on the scalp reduces the spatial coherence of low-level lasers, temporal coherence remains, which may lead to brain damage (Paltsev and Levina 1996). Consequently, it remains uncertain whether the safety findings of LED-based studies apply to laser-based applications. Furthermore, the unique physical properties of the NIR-II region, such as lower refraction, reflection, scattering, and absorption by melanin (Lin et al. 2024), and its specific effects on cognitive performance (Zhao et al. 2022) suggest that it is inappropriate to generalize the safety hazards identified with NIR-I to NIR-II applications.

With the increasing utilization of tPBM in the NIR-II region, particularly at wavelengths of 1060-1070 nm, research in this field lacks comprehensive safety studies to bridge the gap between standardized objective quantification and clinical applicability. Therefore, our study conducted thirteen experiments to robustly explore the safety and protection parameters of tPBM, aimed to provide reliable and quantifiable safety evidence.

## Results

**NSE decrease represents neuroprotection**

In Experiment 1, to evaluate potential tPBM-induced neuronal and glial injuries, we measured NSE and S100β at three time points for each participant: Pre (before tPBM), Post1 (after initiation), and Post2 (60 min post-treatment). As shown in Figure 1, a significant decrease in NSE concentration was observed at both Post1 ($p = 0.002$, Cohen's d = 0.641) and Post2 ($p < 0.001$, Cohen's d = 0.877) relative to the pre-measurements. No substantial changes were observed between the two post-measurement points (Post1 vs. Post2; $ps > 0.451$).

With regards to the S100β concentration levels, no significant differences were observed across the time points ($ps > 0.111$). Furthermore, no instances of detected values exceeded safety standards at any time point, with all values remaining below established safety thresholds (NSE < 20.000 ng/mL; S100β < 0.200 ng/mL).

## Voxel-based morphometry analysis showed no gray matter damage

In experiment 2, T1-weighted (T1w) imaging, which is known for its high spatial resolution, can effectively differentiate between the gray matter, white matter, and cerebrospinal fluid in the brain, making it particularly suitable for detailed exploration of structural changes (Chaieb *et al.* 2014). To ensure that tPBM did not cause structural damage to the brain, we compared structural T1w images taken at Pre (before tPBM) and Post (after 8 min tPBM) sessions of the same participant. We employed voxel-based morphometry (VBM) to investigate voxel-wise differences in local gray matter volume (Douaud et al. 2007), with the aim of eliminating potential safety hazards regarding cortical structures. The results of the permutation-based non-parametric testing revealed no significant differences near the stimulus site (*ps* > 0.010). Visual inspection of T1w images confirmed the absence of any changes (Figure 2). These findings suggested that tPBM did not cause structural damage to the brain.

## No clinical EEG markers suspicious of epileptic discharge were detected

In experiment 3, resting EEG data were collected from participants at Pre (before tPBM) and Post (after 8 min tPBM) sessions to assess potential epileptogenic effects, specifically the emergence of high-frequency neural oscillations, a known indicator of induced epilepsy. We conducted power spectrum density (PSD) analysis at the whole-brain level. By averaging the neural oscillations across all brain regions, the paired t-test results showed no significant differences between Pre and Post sessions across all frequency bands, including delta, theta, alpha, beta, and gamma (*ps* > 0.342). These findings suggested that tPBM does not induce epilepsy. Furthermore, no clinical EEG markers suspicious of epileptic discharge were detected by a certified clinical EEG specialist (Figure 3).

## Questionnaires

In experiment 4, the modified questionnaire developed by Fertonani et al. was used to collect data on subjective feelings of the participants and exclude some negative experiences (Fertonani *et al.* 2010, 2015). Using this questionnaire, we collected subjective reports from the participants after they received sham or active tPBM. Sham sessions employed an active control method to mitigate expectation effects and subjective biases. Our findings revealed that the participants were unable to accurately determine whether they had received the sham or active tPBM session, with their responses aligned with random guess levels at an accuracy rate of 50%, which is consistent with the findings of previous research (Zhao *et al.* 2022).

As illustrated in Figure 4, the average scores for warmth and drowsiness were approximately 2, whereas the scores for the other five dimensions (fatigue, itching, pain, burning, and dizziness) remained below 1, indicating that these sensations were minimally perceived. Warmth, drowsiness, and fatigue were the most commonly reported symptoms among the seven types of discomfort, as illustrated in Figure 4. Following both the active and sham tPBM sessions, all participants reported experiencing warmth. The proportions of individuals reporting drowsiness were high at 90% and 100%, and those reporting fatigue were 86.67% and 90%, respectively. Moreover, there were no significant differences in the ratings of the seven discomfort dimensions between the sham and active tPBM sessions, suggesting that active tPBM did not induce any perceptible discomfort in the participants' subjective experiences.

## Executive function

Experiments 5-13 were conducted to assess potential cognitive impairment by evaluating general executive function. The executive function comprises of three subcomponents: updating, inhibition, and flexibility (Miyake *et al.* 2000). Updating is evaluated through the 2-back task, where participants must identify whether the current stimulus matches that from two prior trials (see Methods for details). Inhibition was measured using the Stroop task, which required participants to resist surrounding interference and make accurate judgments. Cognitive flexibility was assessed by using a switching task. In the initial two non-switch blocks of the switch task, the participants only needed to complete a single task, whereas in the final switch block, they alternated between the two tasks randomly (See the Methods section for details). Executive functions cannot be measured by a single cognitive task, but requires tasks that cover three subcomponents – updating, inhibition, and cognitive flexibility including materials for both cool and hot executive functions (Zelazo and Carlson 2012)

The dependent indicator for updating is the d-prime (d'), where a higher d' value indicates that the target is more easily detected, reflecting a stronger updating ability

(Pelegrina *et al.* 2015; Swets *et al.* 1961). In the Stroop task, inhibition is quantified by calculating the difference in reaction time between conflict and non-conflict conditions; a smaller difference indicates a stronger ability to resist interference (Stroop 1935). For cognitive flexibility, the difference in reaction times between the switch and non-switch blocks indicates the cognitive cost of multitasking (Bialystok 2010). A lower cognitive cost implies a greater capacity for cognitive flexibility.

As shown in Figure 5, the results from the independent sample t-test between the two participant groups (18 for the sham group and 19 for the active group) indicated no significant differences in any of the subcomponents dependent indicators ($ps > 0.090$). This suggests that active tPBM does not impair executive function.

## Discussion

Our study employed multidimensional and quantitative methods to investigate safety concerns associated with tPBM using a low-level laser in the NIR-II region. Thirteen experiments were conducted including blood sample collection, EEG testing, structural MRI scans, questionnaire responses, and executive function tests. The analysis calculated NSE concentration, S100β levels, EEG oscillations, structural morphology, subjective report scores, and executive function capabilities. Our research utilized various designs to ensure the reliability of the safety studies. For the blood sample, EEG assessments, and MRI, a within-subject pre- and post-design was selected to monitor changes in individual physiological indicators caused by tPBM. For the feeling's questionnaires, a within-participants design comparing the active and sham sessions was used to ensure consistent judgment criteria among participants. A between-participants design was implemented for the executive function tests to mitigate the confusion arising from practice effects. Our findings indicated no significant dysfunction in S100β concentration, epileptic EEG markers, gray matter volume, executive capabilities, and subjective sensations attributed to tPBM. Interestingly, we observed a significant reduction in NSE concentration, suggesting the neuroprotective effects of tPBM (Barolet and Boucher 2010). These results provide strong evidence that the application of low-level laser stimulation in the NIR-II region is both safe and potentially beneficial for brain health.

Clinical studies have already demonstrated that tPBM in the NIR-II region offers protective effects on non-brain tissues (Pekyavas and Baltaci 2016; Yesil *et al.* 2020; Yilmaz *et al.* 2022) effectively alleviating pain in conditions, such as knee arthropathies and spinal, shoulder, or elbow disorders, and aiding wound healing (Penberthy and Vorwaller 2021). Our findings expand these observations and reveal a significant decrease in NSE concentration following tPBM, which has also been reported in studies involving mice (Tsai *et al.* 2022). Notably, in this study, the reduced NSE levels persisted for at least 1 h, and tPBM was administered at a wavelength of 1064 nm. NSE and S100β in serum are widely recognized as indicators of neuronal damage (Persson *et al.* 2018). The observed reduction in NSE levels may be linked to the inhibition of neuroinflammation (Barolet and Boucher 2010) and astrocyte proliferation (Massri *et al.* 2018). These findings suggest that tPBM not only possesses a safe profile but also exhibits neuroprotective potential against neuronal damage. No significant reduction in NSE has been reported (Nitsche *et al.* 2003; Oliviero *et al.* 2015; Ullrich *et al.* 2013) with other non-invasive brain stimulations, such as transcranial direct current stimulation, transcranial static magnetic field stimulation, and repetitive transcranial magnetic stimulation (rTMS). To the best of our knowledge, our study is the first to document the impact of tPBM on neurological biomarkers in human serum.

Neuroimaging technologies, including MRI and EEG, have been used to assess the effect of tPBM on brain structure and activity. These tools help ensure that there is no gray matter damage or increased risk of epilepsy, which is a concern with noninvasive brain stimulation due to their modulation of cortical excitability. Excessive excitability can also increase the risk of seizures (Faiman *et al.* 2021; Pegg *et al.* 2020). However, the effects of tPBM on cortical excitability are not well understood. One study observed increased corticospinal excitability following tPBM using single-pulse recordings of motor-evoked potentials (Song *et al.* 2020). Conversely, other studies showed decreased motor evoked potentials after tPBM (Chaieb *et al.* 2015; Konstantinović *et al.* 2013), reflecting decreased corticospinal excitability. Given these mixed results and the ongoing uncertainty regarding the impact on cortical excitability (Dole *et al.* 2023), it was necessary to employ EEG to rule out seizure induction in this study. Thus, EEG is used to detect functional abnormalities, such as epilepsy, and MRI is used to assess structural integrity. The VBM results were also consistent with other noninvasive brain stimulation results (Chaieb *et al.* 2015), showing no gray matter structural changes near the stimulation target. This finding was expected, as the energy emitted by the laser in our study was far below the maximum permissible exposure of the skin. These results provide quantitative evidence to rule out potential risks associated with structural damage or abnormal neuronal discharge.

Several studies have investigated the beneficial effects of tPBM on cognitive function (Lee *et al.* 2023). A review of 35 studies found that 31 reported improvements in at least one domain of cognition. For example, Zhao et al. found that tPBM applied to the right prefrontal cortex improved the visual working memory capacity (Zhao *et al.* 2022). Few studies have reported unchanged or negative effects. We were interested in executive function because to date, tPBM has predominantly targeted the prefrontal cortex, which is closely associated with executive function (Alvarez and Emory 2006). Given that multiple components of executive function are functionally separated within the frontal lobe (Friedman and Robbins 2022), we designed our study to explore aspects of executive function, including inhibition, updating, and flexibility. To the best of our knowledge, this is the first comprehensive evaluation of the safety of tPBM on executive function. Our findings indicated no observed change in executive function, which may be because we only stimulated the prefrontal cortex once, and subsequent studies should use multiple tPBM sessions to improve cognitive function.

For tolerance, the participants reported no significant differences between the active and sham sessions in the following dimensions: itching, pain, burning, warmth, dizziness, drowsiness, and fatigue. Generally, noninvasive brain stimulation is often accompanied by noise and strong skin sensations. For example, the delivery of the magnetic pulse of TMS is associated with a loud clicking sound that may exceed 110 dB (Starck *et al.* 2009), whereas transcranial electrical stimulation may cause sensations of itching or pain (Fertonani *et al.* 2015). These sensations can reduce participants' cooperation and tolerability. However, our results demonstrated that participants undergoing tPBM reported little itching or pain, with no significant differences between the active and sham tPBM sessions. Moreover, tPBM itself did not generate any noticeable noise, and the participants described their sensations during tPBM as mild and barely perceptible. In terms of warmth, previous studies have reported that tPBM can influence skin temperature (Morries *et al.* 2015), which is consistent with our results, which showed a sensation of warmth but not burning during both the sham and active sessions. Interestingly, we observed relatively high drowsiness scores during tPBM stimulation. One plausible explanation is the quiet environment and closed-eye state during tPBM, coupled with the absence of pain-related sensations. These characteristics make tPBM particularly appealing for clinical use in populations with low tolerance, including pediatric patients (Lieske *et al.* 2023).

Our conclusions are limited to the effects observed at specific wavelengths, power densities, and exposure durations. tPBM interventions possess distinct parameters compared to other noninvasive brain stimulation methods, such as irradiance, stimulation duration, wavelength, and light sources (Dole *et al.* 2023). To enhance the comprehensiveness of the safety conclusions, future research should expand the scope of examination to include a broader range of these parameters. Additionally, while brain oscillations extracted from resting-state EEG are useful for diagnosing conditions such as secondary epilepsy, short-term EEG monitoring may not completely exclude this condition, as the induction of epileptic seizures often necessitates prolonged observation (Pegg *et al.*

2020). We also found that tPBM tended to induce drowsiness. Future studies may need to conduct more sophisticated sleep scales to rule out potential sleep-related risks.

Our multidimensional findings, combined with the existing knowledge, enhance our understanding of the safety profiles of tPBM. This study represents an initial step toward advancing clinical utility of tPBM in diverse patient populations. Future studies should be conducted in which precision medicine leverages multimodal imaging to investigate tailored targets, personalized parameters, and their biophysical and neurobiological underpinnings.

## Materials and Methods

### Participants

Thirteen experiments were conducted, and 125 healthy participants were recruited. In Experiment 1, 15 participants (7 men; mean age 27.4 years; age range 22.3–43.8 years) underwent NSE and S100β measurements. In Experiment 2, 15 participants (2 men; mean age 23.4 years; age range 18.4–30.3 years) underwent MRI recording. In Experiment 3, 28 participants (2 men; mean age 24.7 years; age range 19.4–28.9 years) received EEG recordings. In Experiment 4, 30 participants (6 men; mean age 23.1 years; age range 17.8–29.7 years) completed a subjective questionnaire. In Experiments 5-13, 37 participants (6 men; mean age 24.7 years; age range 17.8–29.7 years) completed an executive function tests.

All participants were informed about all aspects of the study and provided written informed consent. None of the participants had any neurological or psychiatric disorders, metallic implants/implanted electric devices, or took any relevant medications at the time of the study. The Institutional Review Board of the local institutes approved the experimental procedures and informed consent was obtained from each participant.

## tPBM protocol

All the experiments were performed using a diode-pumped solid-state laser with a linewidth of approximately ±1 nm (Model JL-LS-100; Jielian Medical Device Inc., Jiangxi, China). The measured uniform laser beam has an area of 13.57 cm$^2$ (4 cm in diameter) and a continuous power output of 2271 mW, resulting in an irradiance or power density of 167 mW/cm$^2$ (2271 mW/13.57 cm$^2$ = 167 mW/cm$^2$) in all experiments except MRI experiment. In the MRI experiment, the measured uniform laser beam has an area of 3.14 cm$^2$ (2 cm in diameter) and a continuous power output of 628 mW, resulting in an irradiance or power density of 200 mW/cm$^2$ (628 mW/3.14 cm$^2$ = 200 mW/cm$^2$). At those power levels, the energy emitted by the laser is far below the maximum permissible exposure for the skin, causing neither detectable physical damage nor imperceptible heat.

In the MRI experiment, a MRI-compatible beamer and headgear were used. The distal end of the optical fiber was equipped with a plastic beamer attached to a custom injection-molded rubber headgear worn by the participants. The headgear was fastened to the participant's head to ensure that its opening was flushed against the forehead. In blood sampling, EEG, and questionnaire experiments, stimulation was administered using a handheld device. In addition, the stimulation site in all experiments was centered on the FP2 electrode, following the 10-20 system used for EEG electrode placement.

During the active tPBM session, a 1064 nm laser was administered for 8 min. Conversely, in the sham tPBM session, two brief 0.5 min stimulations were applied, one at the beginning and another at the end of 8 min. These stimulations were separated by a 7 min interval during which no stimulation was provided, and the laser power was adjusted to 0 W. As a result, the sham tPBM session received approximately one-eighth of the cumulative energy density of the active session. The 0.5 min treatment is crucial in active placebo treatments, as it provides a subjective experience similar to an active tPBM session, although not eliciting any physiological or cognitive effects.

## Blood samples

In experiment 1, to assess the safety in terms of brain damage, we measured serum NSE and S100β, which are indicative of neuronal and glial damage (Persson *et al.* 2018), respectively. Blood samples were collected from 15 participants at 3 time points: Pre (before tPBM), Post1 (after initiation), and Post2 (60 min post-treatment).

Three blood samples were drawn from the antecubital vein with the left and right sides randomly selected. The participants were instructed to rest quietly for 10 min before each draw. After collection, blood samples were immediately centrifuged at 4000 rpm for 5 min. Following centrifugation, the serum was aspirated and stored in a refrigerator at 4 °C. The testing was completed on the same day. The serum NSE levels were measured using an NSE assay kit (Chemiluminescent Magnetic Microparticle Method). S100β quantification was performed using the S100β Quantitative Rapid Test (Lateral Flow Immunoassay). In both cases, adherence to the manufacturer's guidelines was maintained. Paired samples t-tests (two-tailed) were performed to compare NSE and S100β values before and after tPBM stimulation.

## Structural MRI

In experiment 2, T1-weighted images were obtained using 3T MRI (Magnetom TIM Trio, Siemens Healthcare, Erlangen, Germany) at Pre (before tPBM) and Post (after the tPBM) sessions for all 15 participants.

T1-weighted images were obtained using a magnetization-prepared rapid gradient echo (MPRAGE) sequence with a sagittal acquisition, with pre session resolution = 0.8 × 0.8 × 0.8 mm$^3$, repetition time (TR) = 2400 ms, and echo time (TE) = 2.22 ms, and post session resolution = 1 × 1 × 1 mm$^3$, TR = 2530 ms, and TE = 2.98 ms.

VBM was performed with FMRIB Software Library (FSL) on T1-weighted images (Smith *et al.* 2004), using the optimized protocol implemented in FSL-VBM (Douaud *et al.* 2007).

Brain-extracted images were segmented into tissue classes (gray matter, white matter, and cerebrospinal fluid) and then nonlinearly registered to the MNI 152 space using FNIRT. A study-specific left-right symmetric gray matter template was created by averaging and flipping the resulting images along the x-axis. The original subject space gray matter segmentations were registered nonlinearly to match this template.

The registered images were then smoothened using a 3 mm isotropic Gaussian kernel. Permutation-based non-parametric testing (5000 permutations, FWE-corrected $P = 0.01$ using threshold-free cluster enhancement), conducted via the 'randomize' function from FSL (Winkler *et al.* 2014), applied a two-sample paired t-test on VBM analysis maps from FSL.

**EEG test**

In experiment 3, resting-state EEG of 28 participants was recorded for 8 min both at Pre (before tPBM) and Post (after tPBM) sessions, using a SynAmps EEG amplifier (Compumedics NeuroScan, USA) and the Curry 8.0 package (Compumedics NeuroScan, USA). This was performed using a Quickcap fitted with 64 silver chloride electrodes arranged in the international 10-20 system. To track eye movements and blinks, we placed two electrodes 1 cm above and below the left eye for vertical movements, and used electrodes at the outer canthus of each eye for horizontal movements. All electrodes were referenced to the left mastoid except for the eye movement sensors, and we ensured that impedance stayed below 5 kΩ. The EEG signals were amplified within the 0.01 to 200 Hz range and digitized online at a sampling rate of 500 Hz.

Resting-state EEG can detect epilepsy-related abnormal discharges (Faiman *et al.* 2021; Pegg *et al.* 2020) such as spikes, sharp waves, and spike-and-wave complexes. We further calculated PSD using a Fast Fourier Transformation and analyzed for both active and sham sessions across EEG frequency bands: delta (1–4 Hz), theta (4.5–7.5 Hz), alpha (8–12 Hz), beta (12.5–30 Hz) and gamma (35–75 Hz). By focusing on these specific EEG frequency bands, we can assess the risk and presence of epilepsy after tPBM (Faiman et al., 2021). Theta frequency oscillations have shown relevance in idiopathic epilepsy and are potential diagnostic markers (Pegg., et al., 2020). High-frequency oscillations, in particular, provide a robust marker for epileptic activity. Two-tailed t-tests (paired samples, critical P-value 0.05) were performed to compare PSD between Pre and Post tPBM sessions.

## Questionnaire

In experiment 4, all 30 participants revived both the active and sham tPBM sessions, with the session order counterbalanced. At the end of each session, the participants were instructed to complete a published subjective sensation questionnaire (Fertonani *et al.* 2010) that included seven types of discomfort: fatigue, itching, pain, burning, warmth, dizziness, and drowsiness, as well as other sources of discomfort or problems.

Each type of discomfort comprised 5 levels: None, Mild, Moderate, Considerable, and Strong, corresponding to scores of 0–4, respectively. "None" indicated not feeling the described sensation, "Mild" indicated a slight sensation, "Moderate" indicated feeling the described sensation, "Considerable" indicated feeling the described sensation to a significant degree, and "Strong" indicated strongly feeling the described sensation. Additionally, the participants were required to report when the discomfort started, how long it lasted, and the extent to which these sensations affected their performance. They were also required to specify whether these sensations were localized to the head or were present at different locations.

After finishing the last session, the participants were asked to judge whether the tPBM sessions they underwent were active or sham sessions.

## Executive Function

In experiments 5-13, we implemented a between-participants design and recruited two groups of participants to overcome potential practice effects. Both the sham (n=18) and active (n=19) groups were required to complete nine tasks for the executive function tests.

Executive function comprises three subcomponents: updating, inhibition, and cognitive flexibility (Miyake *et al.* 2000). Updating involves manipulating working memory, inhibition entails overcoming interfering cues, and cognitive flexibility assesses multitasking abilities. Tasks in three modalities, emotional, spatial, and letter, were conducted for each subcomponent by controlling material types (Friedman and Robbins 2022), across a total of 9 experiments.

Each task comprised practice and formal blocks. If the accuracy rate in the practice session reached 75% or after completing three practice sessions, the participant could enter the formal session. During each trial, a stimulus image was presented for 3 s, followed by a 0.5 s fixation cross. The participants were required to make accurate and quick judgments based on the appearance of the image according to the rules of each task.

If the accuracy rate was below 70% for a task, the task data from these participants were excluded. Additionally, all incorrect responses, responses with reaction times (RTs) shorter than 120 ms, or responses with a latency exceeding the mean by more than 2.5 standard deviations (for each participant and task separately) were excluded from the RT analyses.

*N-back Task*

Updating was assessed using the 2-back task, in which participants needed to determine if the current material matched the one presented two trials earlier. If it matched, it was considered a target trial; otherwise, it was considered a non-target trial. The 2-back task consisted of 72 trials, 25% of which were target trials.

We chose the 2-back task because it can effectively measure the updating component and has an appropriate difficulty level (Ciesielski *et al.* 2006). The mean of the RTs for correctly identified targets and the individual discrimination indices (d') for performance accuracy were used as dependent variables, with the d' calculated based on the signal

detection theory as follows:

$$d' = z\,(hits/number\ of\ targets) - z\,(false\ alarms/number\ of\ distractors).$$

The resulting value range was $-4.66 \leqslant d' \leqslant +4.66$, with more negative values indicating poorer performance.

For the letter 2-back task, stimuli were numbers 1, 2, 3, 4, 6, 7, 8, and 9. For the spatial 2-back task, the visual stimuli were blue squares presented randomly in one of the eight locations around a fixation (Ragland *et al.* 2002). The emotional 2-back task used facial-expression images of Asian women from the NimStim facial-expression dataset (Tottenham *et al.* 2009).

*Stroop Task*

Inhibition was evaluated using the Stroop task (Stroop 1935), in which participants needed to overcome interference to make correct judgments. The Stroop task consisted of 72 trials, 25% of which featured conflicting elements. Stroop-effect is quantified by calculating the difference in reaction time between conflict and non-conflict conditions. A smaller stroop-effect indicates a stronger ability to resist interference (Stroop 1935).

The materials for the letter Stroop task consisted of four Chinese letters: "红" (red), "绿" (green), "黄" (yellow), and "蓝" (blue), with each letter randomly displayed in one of these four colors. The participants were instructed to identify the font color while inhibiting interference from the meaning of the letters.

For the spatial Stroop task, the materials consisted of four arrows positioned in four cardinal directions around a central fixation cross. The four arrows randomly pointed up, down, left, and right. The participants were instructed to identify the direction indicated by the arrow while inhibiting interference from the cardinal directions around the central fixation cross.

To select emotional images for the Stroop task, we chose materials depicting Asian men and women displaying four emotions: laughter, smile, anger, and sadness. Chinese emotional words were displayed at the center of each face as interference (Tottenham *et al.*

2009). The participants were instructed to identify the emotions depicted in the images.

*Switch Task*

Cognitive flexibility was measured using a switch task (Altgassen *et al.* 2014; Miyake *et al.* 2000), in which participants completed tasks across three blocks: two non-switch blocks with single task and one switch block with dual tasks. The border color of the stimulus images indicates the task category, remaining constant in the non-switch blocks and changing pseudo-randomly in the switch block. Each non-switch block consisted of 64 trials, whereas the switch block consisted of 128 trials. The difference in reaction times between the switch and non-switch blocks indicates switch effect or the cognitive cost of multitasking (Altgassen *et al.* 2014; Miyake *et al.* 2000). A lower switch effect implies a greater capacity for cognitive flexibility.

For the letter switch task, the stimulus materials included the numbers 1, 2, 3, 4, 6, 7, 8, and 9. When the border was red, participants were instructed to determine whether the number was greater or less than five. When the border was blue, the participants were instructed to determine whether the number was odd or even.

The stimuli for the spatial switch task encompassed 16 images, each exhibiting one of four large patterns created through the integration of small shapes, such as circles, squares, triangles, and hearts. When the border was red, participants performed a small-pattern judgment task. When the border was blue, the participants performed the large pattern judgment task.

Emotional images were chosen as material for the emotional-switch task, and the stimuli were images of Asian men and women expressing four emotions: laughter, smile, anger, and sadness. When the border was red, participants were instructed to perform a sex judgment task. When the border was blue, participants were directed to perform a smile judgment task.

# Figure Legends

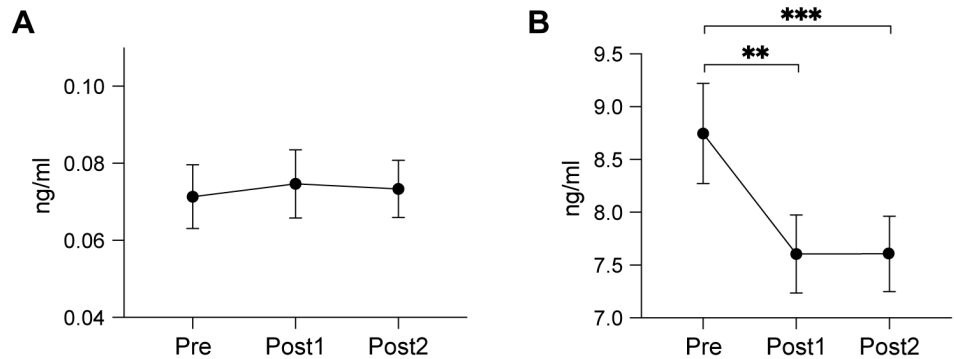

**Figure 1. Changes in concentration levels of S100β (A) and NSE (B) at three time points (n = 15).** All measured indicators remained below the safety threshold. For NSE, concentrations are significantly reduced following tPBM, and this reduction is sustained up to 1 h post-treatment. In contrast, no significant changes in concentration are observed for S100β. NSE, neuron specific enolase; tPBM, transcranial photobiomodulation.

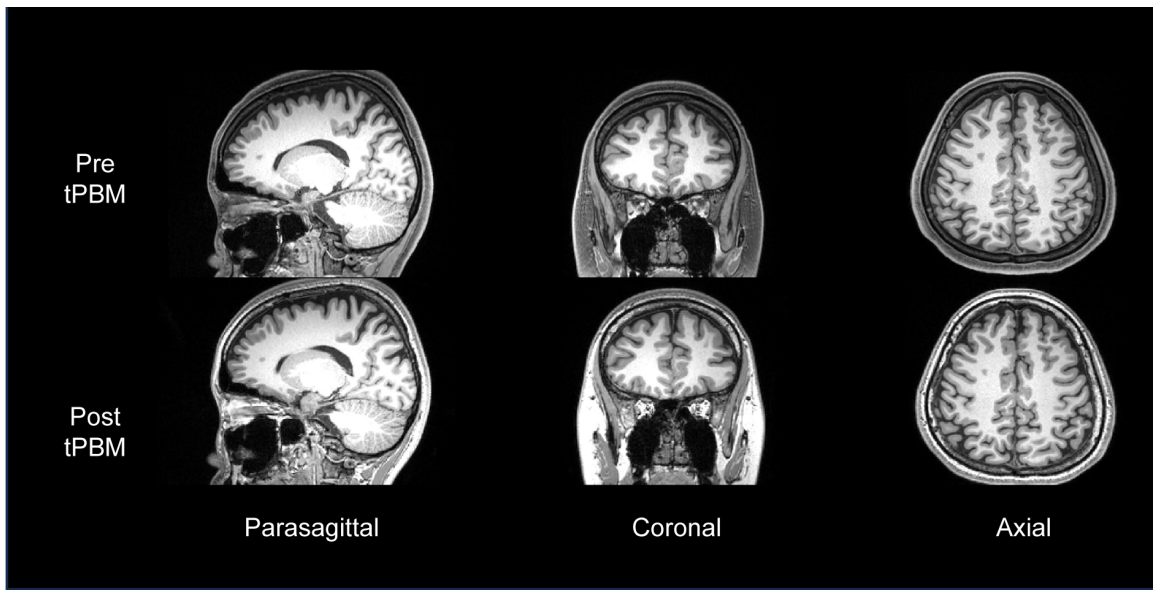

**Figure 2. Pre and Post tPBM MRI T1-weighted images (n = 15).** MRI T1w images were taken Pre (before tPBM) and Post (after 8 min of tPBM) sessions. One of the representative T1w images is shown here. VBM analysis shows no significant change of gray matter volume at the stimulation site. tPBM, transcranial Photobiomodulation; VBM, voxel-based morphometry.

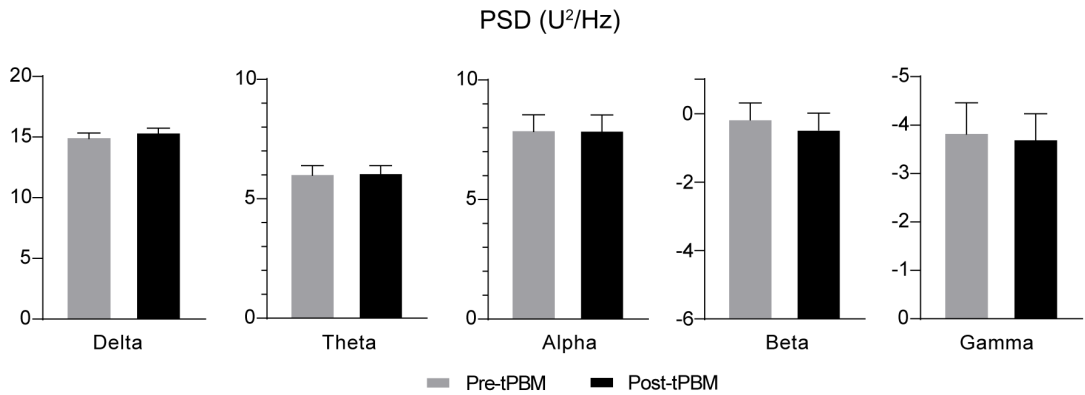

**Figure 3. Comparison of brain oscillations Pre and Post tPBM sessions (n = 28)**. The oscillatory activity across the delta, theta, alpha, beta, and gamma frequency ranges. No significant differences are observed between the Pre and Post tPBM sessions in any oscillatory activity. tPBM, transcranial photobiomodulation.

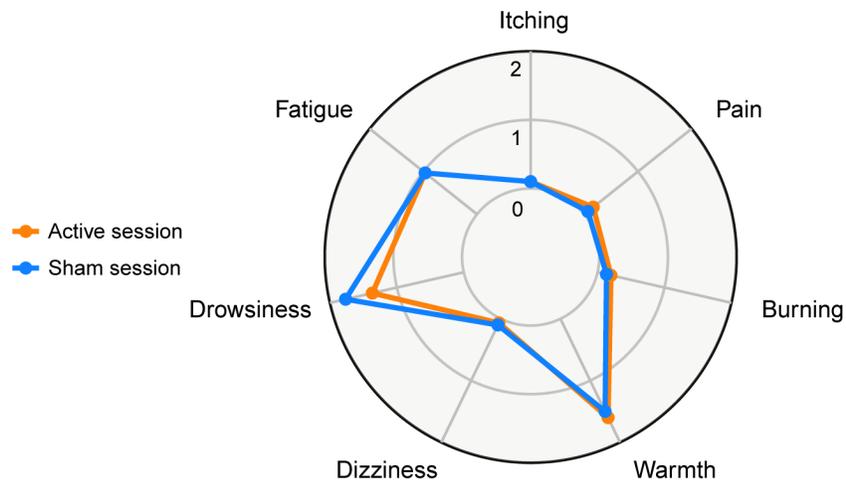

**Figure 4. Average scores from the tPBM subjective feelings questionnaire (n=30).** Scores of 0, 1 and 2 represent none, mild and moderate sensations, respectively. Except for warmth and drowsiness, the average scores for the remaining five dimensions were below 2, indicating very weak sensations. No statistically significant differences were observed across all dimensions between sham and active tPBM sessions. tPBM, transcranial photobiomodulation.

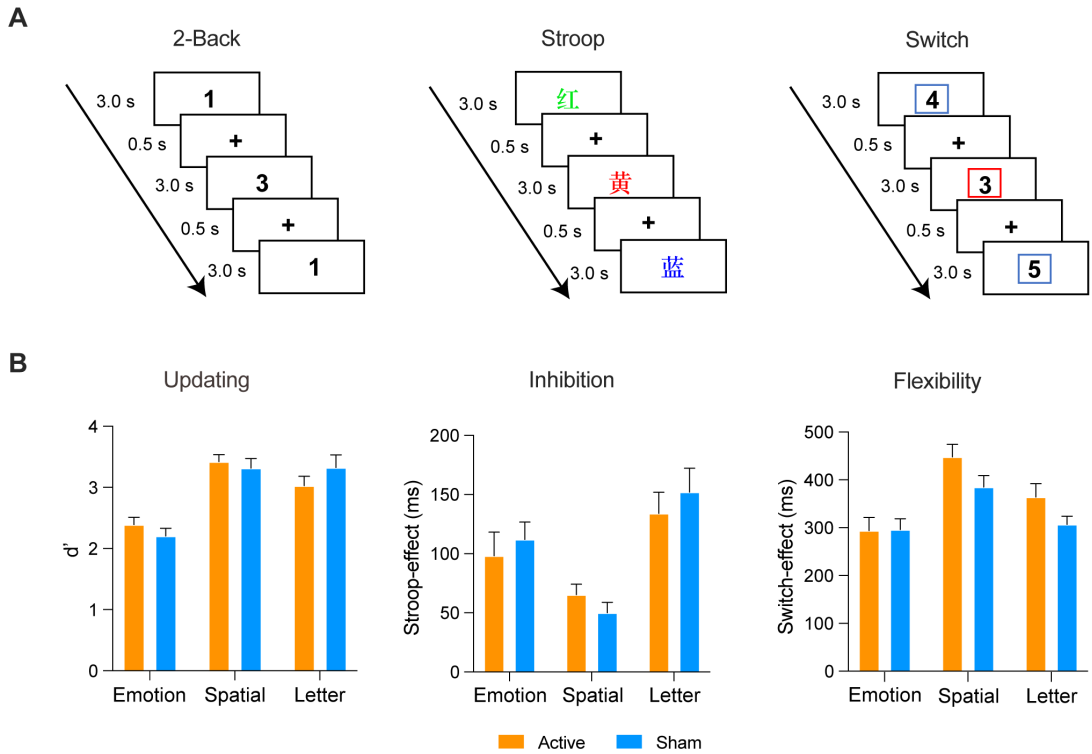

**Figure 5. Dependency index of the three subcomponents of executive function (n=19 for active group, n=18 for sham group).** (A) The measurement of the executive function contains three tasks: N-back, Stroop, and Switch with each subcomponent containing three modalities: emotion, space, and letter. The letter modality is shown in the figure. (B) No differences are found between the active and sham participants groups at all dependency indexes.

# References

Altgassen M, Vetter NC, Phillips LH, Akgün C and Kliegel M (2014) Theory of mind and switching predict prospective memory performance in adolescents. *Journal of Experimental Child Psychology* 127, 163–175. https://doi.org/10.1016/j.jecp.2014.03.009.

Alvarez JA and Emory E (2006) Executive Function and the Frontal Lobes: A Meta-Analytic Review. *Neuropsychology Review* 16(1), 17–42. https://doi.org/10.1007/s11065-006-9002-x.

Barolet D and Boucher A (2010) Radiant near infrared light emitting Diode exposure as skin preparation to enhance photodynamic therapy inflammatory type acne treatment outcome. *Lasers in Surgery and Medicine* 42(2), 171–178. https://doi.org/10.1002/lsm.20886.

Bialystok E (2010) Global–Local and Trail-Making Tasks by Monolingual and Bilingual Children: Beyond Inhibition. *Developmental Psychology* 46(1), 93–105. https://doi.org/10.1037/a0015466.

Caldieraro MA and Cassano P (2019) Transcranial and systemic photobiomodulation for major depressive disorder: A systematic review of efficacy, tolerability and biological mechanisms. *Journal of Affective Disorders* 243, 262–273. https://doi.org/10.1016/j.jad.2018.09.048.

Cassano P, Norton R, Caldieraro MA, Vahedifard F, Vizcaino F, McEachern KM and Iosifescu D (2022) Tolerability and Safety of Transcranial Photobiomodulation for Mood and Anxiety Disorders. *Photonics* 9(8), 507. https://doi.org/10.3390/photonics9080507.

Chaieb L, Antal A, Masurat F and Paulus W (2015) Neuroplastic effects of transcranial near-infrared stimulation (tNIRS) on the motor cortex.

*Frontiers in Behavioral Neuroscience* 9, 147.

https://doi.org/10.3389/fnbeh.2015.00147.

Chaieb L, Antal A, Pisoni A, Saiote C, Opitz A, Ambrus GG, Focke N and Paulus W (2014) Safety of 5 kHz tACS. *Brain Stimulation* 7(1), 92–96. https://doi.org/10.1016/j.brs.2013.08.004.

Ciesielski KT, Lesnik PG, Savoy RL, Grant EP and Ahlfors SP (2006) Developmental neural networks in children performing a Categorical N-Back Task. *NeuroImage* 33(3), 980–990.

https://doi.org/10.1016/j.neuroimage.2006.07.028.

Cotler HB, Chow RT, Hamblin MR and Carroll J (2015) The Use of Low Level Laser Therapy (LLLT) For Musculoskeletal Pain. *MOJ Orthopedics & Rheumatology* 2(5). https://doi.org/10.15406/mojor.2015.02.00068.

Dole M, Auboiroux V, Langar L and Mitrofanis J (2023) A systematic review of the effects of transcranial photobiomodulation on brain activity in humans. *Reviews in the Neurosciences* 34(6), 671–693.

https://doi.org/10.1515/revneuro-2023-0003.

Douaud G, Smith S, Jenkinson M, Behrens T, Johansen-Berg H, Vickers J, James S, Voets N, Watkins K, Matthews PM and James A (2007) Anatomically related grey and white matter abnormalities in adolescent-onset schizophrenia. *Brain* 130(9), 2375–2386.

https://doi.org/10.1093/brain/awm184.

Faiman I, Smith S, Hodsoll J, Young AH and Shotbolt P (2021) Resting-state EEG for the diagnosis of idiopathic epilepsy and psychogenic nonepileptic seizures: A systematic review. *Epilepsy & Behavior* 121(Pt A), 108047. https://doi.org/10.1016/j.yebeh.2021.108047.

Fertonani A, Ferrari C and Miniussi C (2015) What do you feel if I apply transcranial electric stimulation? Safety, sensations and secondary

induced effects. *Clinical Neurophysiology* 126(11), 2181–2188. https://doi.org/10.1016/j.clinph.2015.03.015.

Fertonani A, Rosini S, Cotelli M, Rossini PM and Miniussi C (2010) Naming facilitation induced by transcranial direct current stimulation. *Behavioural Brain Research* 208(2), 311–318. https://doi.org/10.1016/j.bbr.2009.10.030.

Friedman NP and Robbins TW (2022) The role of prefrontal cortex in cognitive control and executive function. *Neuropsychopharmacology* 47(1), 72–89. https://doi.org/10.1038/s41386-021-01132-0.

Gutiérrez-Menéndez A, Marcos-Nistal M, Méndez M and Arias JL (2020) Photobiomodulation as a promising new tool in the management of psychological disorders: A systematic review. *Neuroscience & Biobehavioral Reviews* 119, 242–254. https://doi.org/10.1016/j.neubiorev.2020.10.002.

Hacke W, Schellinger PD, Albers GW, Bornstein NM, Dahlof BL, Fulton R, Kasner SE, Shuaib A, Richieri SP, Dilly SG, Zivin J, Lees KR, Broderick J, Ivanova A, Johnston K, Norrving B, Alexandrov A, Brown D, Capone P, Chiu D, Clark W, Cochran J, Deredyn C, Devlin T, Hickling W, Howell G, Huang D, Hussain S, Mallenbaum S, Moonis M, Nash M, Rymer M, Taylor R, Tremwel M, Buck B, Perez J, Gerloff C, Greiwing B, Grond M, Hamman G, Haarmeiter T, Jander S, Köhrmann M, Ritter M, Schneider D, Sobesky J, Steiner T, Steinmetz H, Veltkamp R, Weimar C, Gruber F, Andersson B, Welin L, Leys D, Tatlisumak T, Luft A, Lyrer P, Michel P, Molina C and Segura T (2014) Transcranial Laser Therapy in Acute Stroke Treatment. *Stroke* 45(11), 3187–3193. https://doi.org/10.1161/strokeaha.114.005795.

Hashmi JT, Huang Y-Y, Osmani BZ, Sharma SK, Naeser MA and Hamblin MR (2010) Role of Low-Level Laser Therapy in Neurorehabilitation. *PM&R* 2(12), S292–S305. https://doi.org/10.1016/j.pmrj.2010.10.013.

Hecht J (2010) Short history of laser development. *Optical Engineering* 49(9), 091002–091002–23. https://doi.org/10.1117/1.3483597.

Hohberger B (2016) Laser and Its Hazard Potential. *Journal of Nuclear Medicine & Radiation Therapy* 7(5), 1–3. https://doi.org/10.4172/2155-9619.1000303.

Huang Y, Nagata K, Tedford CE, McCarthy T and Hamblin MR (2013) Low-level laser therapy (LLLT) reduces oxidative stress in primary cortical neurons in vitro. *Journal of Biophotonics* 6(10), 829–838. https://doi.org/10.1002/jbio.201200157.

Huisa BN, Stemer AB, Walker MG, Rapp K, Meyer BC, Zivin JA and investigators N and -2 (2013) Transcranial Laser Therapy for Acute Ischemic Stroke: A Pooled Analysis of NEST-1 and NEST-2. *International Journal of Stroke* 8(5), 315–320. https://doi.org/10.1111/j.1747-4949.2011.00754.x.

Konstantinović LM, Jelić MB, Jeremić A, Stevanović VB, Milanović SD and Filipović SR (2013) Transcranial application of near-infrared low-level laser can modulate cortical excitability. *Lasers in Surgery and Medicine* 45(10), 648–653. https://doi.org/10.1002/lsm.22190.

Lampl Y, Zivin JA, Fisher M, Lew R, Welin L, Dahlof B, Borenstein P, Andersson B, Perez J, Caparo C, Ilic S and Oron U (2007) Infrared Laser Therapy for Ischemic Stroke: A New Treatment Strategy. *Stroke* 38(6), 1843–1849. https://doi.org/10.1161/strokeaha.106.478230.

Lapchak PA, Wei J and Zivin JA (2004) Transcranial Infrared Laser Therapy Improves Clinical Rating Scores After Embolic Strokes in Rabbits. *Stroke* 35(8), 1985–1988. https://doi.org/10.1161/01.str.0000131808.69640.b7.

Lee T, Ding Z and Chan AS (2023) Can transcranial photobiomodulation improve cognitive function? A systematic review of human studies. *Ageing Research Reviews* 83, 101786. https://doi.org/10.1016/j.arr.2022.101786.

Li T, Wang P, Qiu L, Fang X and Shang Y (2018) Optimize Illumination Parameter of Low-Level Laser Therapy for Hemorrhagic Stroke by Monte Carlo Simulation on Visible Human Dataset. *IEEE Photonics Journal* 10(3), 1–9. https://doi.org/10.1109/jphot.2018.2834740.

Liang HL, Whelan HT, Eells JT, Meng H, Buchmann E, Lerch-Gaggl A and Wong-Riley M (2006) Photobiomodulation partially rescues visual cortical neurons from cyanide-induced apoptosis. *Neuroscience* 139(2), 639–649. https://doi.org/10.1016/j.neuroscience.2005.12.047.

Lieske A, Kellman M, Houlihan K, Hodapp S, Chen M, Jacob S, Lim K and Conelea C (2023) Nibs and kids: a program development project to establish a pediatric-dedicated non-invasive brain stimulation facility. *Brain Stimulation* 16(1), 371. https://doi.org/10.1016/j.brs.2023.01.730.

Lin H, Li D, Zhu J, Liu S, Li J, Yu T, Tuchin VV, Semyachkina-Glushkovskaya O and Zhu D (2024) Transcranial photobiomodulation for brain diseases: review of animal and human studies including mechanisms and emerging trends. *Neurophotonics* 11(01), 010601. https://doi.org/10.1117/1.nph.11.1.010601.

Massri NE, Weinrich TW, Kam JH, Jeffery G and Mitrofanis J (2018) Photobiomodulation reduces gliosis in the basal ganglia of aged mice. *Neurobiology of Aging* 66, 131–137. https://doi.org/10.1016/j.neurobiolaging.2018.02.019.

Miyake A, Friedman NP, Emerson MJ, Witzki AH, Howerter A and Wager TD (2000) The Unity and Diversity of Executive Functions and Their Contributions to Complex "Frontal Lobe" Tasks: A Latent Variable

Analysis. *Cognitive Psychology* 41(1), 49–100. https://doi.org/10.1006/cogp.1999.0734.

Mochizuki-Oda N, Kataoka Y, Cui Y, Yamada H, Heya M and Awazu K (2002) Effects of near-infra-red laser irradiation on adenosine triphosphate and adenosine diphosphate contents of rat brain tissue. *Neuroscience Letters* 323(3), 207–210. https://doi.org/10.1016/s0304-3940(02)00159-3.

Morries LD, Cassano P and Henderson TA (2015) Treatments for traumatic brain injury with emphasis on transcranial near-infrared laser phototherapy. *Neuropsychiatric Disease and Treatment* 11, 2159–2175. https://doi.org/10.2147/ndt.s65809.

Nitsche MA, Nitsche MS, Klein CC, Tergau F, Rothwell JC and Paulus W (2003) Level of action of cathodal DC polarisation induced inhibition of the human motor cortex. *Clinical Neurophysiology* 114(4), 600–604. https://doi.org/10.1016/s1388-2457(02)00412-1.

Oliviero A, Carrasco-López MC, Campolo M, Perez-Borrego YA, Soto-León V, Gonzalez-Rosa JJ, Higuero AM, Strange BA, Abad-Rodriguez J and Foffani G (2015) Safety Study of Transcranial Static Magnetic Field Stimulation (tSMS) of the Human Cortex. *Brain Stimulation* 8(3), 481–485. https://doi.org/10.1016/j.brs.2014.12.002.

Oron A, Oron U, Streeter J, Taboada LD, Alexandrovich A, Trembovler V and Shohami E (2007) Low-Level Laser Therapy Applied Transcranially to Mice following Traumatic Brain Injury Significantly Reduces Long-term Neurological Deficits. *Journal of Neurotrauma* 24(4), 651–656. https://doi.org/10.1089/neu.2006.0198.

Paltsev Yu and Levina A (1996) Biophysical and medical safety basis of laser emission. *CIS Selected Papers: Laser Use in Oncology* 137–147. https://doi.org/10.1117/12.229482.

Pegg EJ, Taylor JR and Mohanraj R (2020) Spectral power of interictal EEG in the diagnosis and prognosis of idiopathic generalized epilepsies. *Epilepsy & Behavior* 112, 107427. https://doi.org/10.1016/j.yebeh.2020.107427.

Pekyavas NO and Baltaci G (2016) Short-term effects of high-intensity laser therapy, manual therapy, and Kinesio taping in patients with subacromial impingement syndrome. *Lasers in Medical Science* 31(6), 1133–1141. https://doi.org/10.1007/s10103-016-1963-2.

Pelegrina S, Lechuga MT, García-Madruga JA, Elosúa MR, Macizo P, Carreiras M, Fuentes LJ and Bajo MT (2015) Normative data on the n-back task for children and young adolescents. *Frontiers in Psychology* 6, 1544. https://doi.org/10.3389/fpsyg.2015.01544.

Penberthy WT and Vorwaller CE (2021) Utilization of the 1064 nm Wavelength in Photobiomodulation: A Systematic Review and Meta-Analysis: *Journal of Lasers in Medical Sciences* 12(1), e86–e86. https://doi.org/10.34172/jlms.2021.86.

Persson L, Hårdemark HG, Gustafsson J, Rundström G, Mendel-Hartvig I, Esscher T and Påhlman S (2018) S-100 protein and neuron-specific enolase in cerebrospinal fluid and serum: markers of cell damage in human central nervous system. *Stroke* 18(5), 911–918. https://doi.org/10.1161/01.str.18.5.911.

Posten W, Wrone DA, Dover JS, Arndt KA, Silapunt S and Alam M (2005) Low‐Level Laser Therapy for Wound Healing: Mechanism and Efficacy. *Dermatologic Surgery* 31(3), 334–340. https://doi.org/10.1111/j.1524-4725.2005.31086.

Pruitt T, Wang X, Wu A, Kallioniemi E, Husain MM and Liu H (2020) Transcranial Photobiomodulation (tPBM) With 1,064‐nm Laser to Improve Cerebral Metabolism of the Human Brain In Vivo. *Lasers in*

Surgery and Medicine 52(9), 807–813. https://doi.org/10.1002/lsm.23232.

Qu X, Li L, Zhou X, Dong Q, Liu H, Liu H, Yang Q, Han Y and Niu H (2022) Repeated transcranial photobiomodulation improves working memory of healthy older adults: behavioral outcomes of poststimulation including a three-week follow-up. *Neurophotonics* 9(3), 035005–035005. https://doi.org/10.1117/1.nph.9.3.035005.

Ragland JD, Turetsky BI, Gur RC, Gunning-Dixon F, Turner T, Schroeder L, Chan R and Gur RE (2002) Working Memory for Complex Figures: An fMRI Comparison of Letter and Fractal n-Back Tasks. *Neuropsychology* 16(3), 370–379. https://doi.org/10.1037/0894-4105.16.3.370.

Rojas JC and Gonzalez-Lima F (2011) Low-level light therapy of the eye and brain. *Eye and Brain* 3, 49–67. https://doi.org/10.2147/eb.s21391.

Salehpour F, Mahmoudi J, Kamari F, Sadigh-Eteghad S, Rasta SH and Hamblin MR (2018) Brain Photobiomodulation Therapy: a Narrative Review. *Molecular Neurobiology* 55(8), 6601–6636. https://doi.org/10.1007/s12035-017-0852-4.

Shahdadian S, Wang X, Wanniarachchi H, Chaudhari A, Truong NCD and Liu H (2022) Neuromodulation of brain power topography and network topology by prefrontal transcranial photobiomodulation. *Journal of Neural Engineering* 19(6), 066013. https://doi.org/10.1088/1741-2552/ac9ede.

Smith SM, Jenkinson M, Woolrich MW, Beckmann CF, Behrens TEJ, Johansen-Berg H, Bannister PR, Luca MD, Drobnjak I, Flitney DE, Niazy RK, Saunders J, Vickers J, Zhang Y, Stefano ND, Brady JM and Matthews PM (2004) Advances in functional and structural MR image analysis and implementation as FSL. *NeuroImage* 23, S208–S219. https://doi.org/10.1016/j.neuroimage.2004.07.051.

Song P, Han T, Lin H, Li S, Huang Q, Dai X, Wang R and Wang Y (2020) Transcranial near-infrared stimulation may increase cortical excitability recorded in humans. *Brain Research Bulletin* 155, 155–158. https://doi.org/10.1016/j.brainresbull.2019.12.007.

Starck J, Rimpiläinen I, Pyykkö I and Esko T (2009) The Noise Level in Magnetic Stimulation. *Scandinavian Audiology* 25(4), 223–226. https://doi.org/10.3109/01050399609074958.

Stroop JR (1935) Studies of interference in serial verbal reactions. *Journal of Experimental Psychology* 18(6), 643–662. https://doi.org/10.1037/h0054651.

Swets JA, Tanner WP and Birdsall TG (1961) Decision Processes In Perception. *Psychological Review* 68(5), 301–340. https://doi.org/10.1037/h0040547.

Tottenham N, Tanaka JW, Leon AC, McCarry T, Nurse M, Hare TA, Marcus DJ, Westerlund A, Casey B and Nelson C (2009) The NimStim set of facial expressions: Judgments from untrained research participants. *Psychiatry Research* 168(3), 242–249. https://doi.org/10.1016/j.psychres.2008.05.006.

Truong NCD, Wang X, Wanniarachchi H and Liu H (2022) Enhancement of Frequency-Specific Hemodynamic Power and Functional Connectivity by Transcranial Photobiomodulation in Healthy Humans. *Frontiers in Neuroscience* 16, 896502. https://doi.org/10.3389/fnins.2022.896502.

Ullrich H, Kranaster L, Sigges E, Andrich J and Sartorius A (2013) Neuron specific enolase and serum remain unaffected by ultra high frequency left prefrontal transcranial magnetic stimulation in patients with depression: a preliminary study. *Journal of Neural Transmission* 120(12), 1733–1736. https://doi.org/10.1007/s00702-013-1050-9.

Uozumi Y, Nawashiro H, Sato S, Kawauchi S, Shima K and Kikuchi M (2010) Targeted increase in cerebral blood flow by transcranial near‐infrared laser irradiation. *Lasers in Surgery and Medicine* 42(6), 566–576. https://doi.org/10.1002/lsm.20938.

Winkler AM, Ridgway GR, Webster MA, Smith SM and Nichols TE (2014) Permutation inference for the general linear model. *NeuroImage* 92(100), 381–397. https://doi.org/10.1016/j.neuroimage.2014.01.060.

Yesil H, Dundar U, Toktas H, Eyvaz N and Yeşil M (2020) The effect of high intensity laser therapy in the management of painful calcaneal spur: a double blind, placebo-controlled study. *Lasers in Medical Science* 35(4), 841–852. https://doi.org/10.1007/s10103-019-02870-w.

Yılmaz M, Eroglu S, Dundar U and Toktas H (2022) The effectiveness of high-intensity laser therapy on pain, range of motion, functional capacity, quality of life, and muscle strength in subacromial impingement syndrome: a 3-month follow-up, double-blinded, randomized, placebo-controlled trial. *Lasers in Medical Science* 37(1), 241–250. https://doi.org/10.1007/s10103-020-03224-7.

Zhao C, Li D, Kong Y, Liu H, Hu Y, Niu H, Jensen O, Li X, Liu H and Song Y (2022) Transcranial photobiomodulation enhances visual working memory capacity in humans. *Science Advances* 8(48), eabq3211. https://doi.org/10.1126/sciadv.abq3211.

Zivin JA, Albers GW, Bornstein N, Chippendale T, Dahlof B, Devlin T, Fisher M, Hacke W, Holt W, Ilic S, Kasner S, Lew R, Nash M, Perez J, Rymer M, Schellinger P, Schneider D, Schwab S, Veltkamp R, Walker M, Streeter J and Investigators NE and ST-2 (2009) Effectiveness and Safety of Transcranial Laser Therapy for Acute Ischemic Stroke. *Stroke* 40(4), 1359–1364. https://doi.org/10.1161/strokeaha.109.547547.